\documentclass[fleqn,usenatbib,usedcolumn]{mnras}
\usepackage{graphicx}	
\usepackage{amsmath}	
\usepackage{amssymb} 
\usepackage{times} 
\usepackage{aas_macros} 
\usepackage[dvipsnames,usenames]{color}

\usepackage{longtable}
\usepackage{array}
\usepackage{soul}

\def\note #1]{\noindent{\bf #1]}}

\def\sgnk2{{\rm sgn\left(K^2\right)}}

\title[Stochastic Signatures]{Modelling Stochastic Signatures in Classical Pulsators} 
\author[Avelino et al.]{P. P. Avelino$^{1,2,3}$, M. S. Cunha$^{1,3}$, W. J. Chaplin$^{3,4}$\\
$^{1}$Instituto de Astrof\'\i sica e Ci\^encias do Espa\c co, Universidade do
Porto, CAUP, Rua das Estrelas, PT4150-762 Porto, Portugal\\
$^{2}$Departamento de Física e Astronomia, Faculdade de Ciências, Universidade do Porto, Rua do Campo Alegre 687, PT4169-007 Porto, Portugal\\
$^3$School of Physics and Astronomy, University of Birmingham, Birmingham, B15 2TT, United Kingdom\\
$^4$Stellar Astrophysics Centre (SAC), Department of Physics and Astronomy, Aarhus University, Ny Munkegade 120, DK-8000 Aarhus C, Denmark\\
}

\date{Accepted XXX. Received YYY; in original form ZZZ}

\pubyear{2019}

\begin{document}
	\label{firstpage}
	\pagerange{\pageref{firstpage}--\pageref{lastpage}}
	\maketitle
\begin{abstract}
{We consider the impact of stochastic perturbations on otherwise coherent oscillations of classical pulsators. The resulting dynamics are modelled by a driven damped harmonic oscillator subject to either an external or an internal forcing and white noise velocity fluctuations. We characterize the phase and relative amplitude variations using analytical and numerical tools.  When the forcing is internal the phase variation displays a random walk behaviour and a red noise power spectrum with a ragged erratic appearance. We determine the dependence of the root mean square phase and relative amplitude variations ($\sigma_{\Delta \varphi}$ and $\sigma_{\Delta A/A}$, respectively) on the amplitude of the stochastic perturbations, the damping constant $\eta$, and the total observation time $t_{\rm obs}$ for this case, under the assumption that the relative amplitude variations remain small, showing that $\sigma_{\Delta \varphi}$ increases with $t_{\rm obs}^{1/2}$ becoming much larger than $\sigma_{\Delta A/A}$ for $t_{\rm obs} \gg \eta^{-1}$. In the case of an external forcing the phase and relative amplitude variations remain of the same order, independent of the observing time. In the case of an internal forcing, we find that $\sigma_{\Delta \varphi}$ does not depend on $\eta$. Hence, the damping time cannot be inferred from fitting the power of the signal, as done for solar-like pulsators, but the amplitude of the stochastic perturbations may be constrained from the observations. Our results imply that, given sufficient time, the variation of the phase associated to the stochastic perturbations in internally driven classical pulsators will become sufficiently large to be probed observationally.}
\end{abstract}


\section{Introduction\label{introduction}}
{Pulsating stars are often classified as either solar-like or classical pulsators, depending on whether their oscillations are intrinsically stable or unstable, respectively ({\it i.e.} on whether small perturbations will, respectively, decay or grow in the linear regime) \cite[e.g.,][]{cunhaetal07,aerts10}.} In solar-like pulsators, the oscillations are stochastically excited by near-surface convection \citep{chaplin13}. As a result of their stochastic nature, their amplitudes and phases vary with time in a manner that is best described in statistical terms. On the other hand, in classical pulsators the excitation follows from the amplification of small disturbances by a coherent forcing, most commonly resulting from the heating of particular layers during compression ({\it e.g.,} due to an increase of the opacity), similar to the process underlying a heat-engine. In classical pulsators the oscillations tend to be stable and coherent on long time scales.

Despite the above, mode frequency (or phase) variability, often accompanied by variations in mode amplitude, have long been observed in classical pulsators \cite[see][for an historical review]{neilson2016}.  Such variations are found across the HR diagram, in giant stars, such as Cepheids \cite[e.g.][]{derekas17,smolec17}, RR Lyrae \cite[e.g.][]{benko19}, and Mira-like stars \cite[e.g.][]{percy1999,bedding05}, compact pulsators, such as white dwarfs \cite[e.g.][]{winget83,winget94,vauclair2011} and subdwarf B stars \cite[e.g.][]{kilkenny10,zong2018}, as well as in a diverse range of main-sequence pulsators, including $\beta$ Cepheid \cite[][]{pigulski08,degroote10}, $\delta$-Scuti \cite[e.g.][]{breger98,bowman2016,breger2017}, and rapidly oscillating Ap stars \cite[e.g.][]{kurtz97,holdworth14,balona2019}. In some cases these variations may have an  extrinsic origin, such as an unseen companion. However, in most cases the origin is likely intrinsic to the star. Common physical mechanisms evoked to explain frequency variability in classical pulsators are secular evolution~\cite[e.g.][]{winget83}, non-linear effects~\cite[e.g.][]{buchler1997,buchler04}, and beating of close frequencies \cite[e.g.][]{breger02}.

An interesting question to address in this context is whether stochastic perturbations, such as those responsible for the driving of solar-like pulsators, will lead to significant phase and amplitude variations in oscillations driven by a coherent forcing.  In fact, such perturbations are expected to be ubiquitous at least in stars with convective unstable envelopes. The idea that random fluctuations could be responsible for period irregularities in classical pulsators, in particular Long Period Variable stars (LPV), dates back to \cite{eddington1929} and has found observational support in many works \cite[e.g.][]{percy1999,benko19}. While the pulsations in LPV stars are generally thought to be intrinsically unstable \citep{trabucchi19}, significant convection-pulsation interaction is expected \cite[e.g.][]{freytag17}, with convection possibly having a significant role in the coherent driving \citep{xiong18}, as well as on the stochastic signature observed in a number of stars \cite[e.g.][]{JCD01,bedding05}.

With the long and ultra-precise photometric time series provided by the {\it Kepler} satellite \citep{borucki10,gillilandetal10} one would expect the impact of stochastic processes on classical pulsators to become more evident across the HR diagram. With this in mind, in this work we introduce a phenomenological model for a coherent pulsator in the presence of a stochastic perturbation based on a modified driven damped harmonic oscillator. We start, in section~\ref{standard}, by describing the impact expected from stochastic noise when the driving of the oscillations is external to the star. We then modify our model to consider the possibility that the phase of the acceleration associated to the driving mechanism varies in time in reaction to the changes to the pulsation phase introduced by the stochastic noise, as expected for internal forcing. The analytical results for the modified model are discussed in section~\ref{modifieda} and the more general numerical results are discussed in section~\ref{modifiedn}. In section~\ref{conclusions} we discuss our findings and conclude.

\section{Externally Driven Damped Harmonic Oscillator with Noise}
\label{standard}
Consider the equation of motion of a driven damped harmonic oscillator with noise as given by 
\label{}
\begin{equation}
{\ddot x}+2 \eta {\dot x} + \omega_0^2 x = a_f + \xi\,. \label{fho}
\end{equation}
Here, a dot represents a derivative with respect to the physical time $t$, $x(t)$ is the displacement, $\omega_0$ is the natural angular frequency of the oscillator, $\eta$ is the damping constant, $a_f$ is the acceleration associated to the driving mechanism, and $\xi(t)$ is the function which parameterizes the noise. {Equation~(\ref{fho}) will be used to describe the dynamics of a single intrinsically unstable mode subject to random fluctuations, such as those produced by convection, here represented by the noise term. In this section we shall consider the case of an external coherent driving, {\it e.g.}, imposed by a companion. The case of an internal driving, {\it e.g.}, resulting from a heat-engine-like mechanism will be discussed in sections~\ref{modifieda} and \ref{modifiedn}. To guarantee that the system in the presence of driving and noise remains in equilibrium, energy must be subtracted in every cycle. That is achieved by assuming that some source of continuous damping is present, parameterised by a characteristic damping time scale of $\eta^{-1}$.}

In the present paper we shall assume that the stochastic perturbations generate random impulses acting on short timescales, which we model as instantaneous finite variations of the velocity, with the displacement being always continuous. We shall consider both the effect of a single instantaneous kick modifying the velocity by $\Delta \dot x$ at an arbitrary time $t_k$, in which case
\begin{equation}
\xi(t) = \Delta \dot x (t_k) \delta(t-t_k)\,,
\end{equation}
where $\delta$ represents the Dirac delta function, or a series of successive random velocity kicks which, for simplicity, are assumed to be separated by a fixed time interval $\Delta t$. In this case $\xi$ may be written as
\begin{equation}
\xi(t) = 2 A_{N} \omega_0     \sqrt{\omega_0 \Delta t} \sum_{k=0}^\infty r(k) \, \delta(t-k\Delta t)\,,
\label{xi}
\end{equation}
where $r(k)$ are independent random variables with a normal distribution of mean zero and unit standard deviation ($r(k) \sim N(0,1)$), and $A_N$, with dimensions of a length, parameterizes the amplitude of the white noise.   As we shall see later, the normalisation in Eq.~(\ref{xi}) is such that the average velocity variation in a timescale equal to the oscillation period $P$ is proportional to $A_N \omega_0$. We shall consider that $\Delta t$ is significantly shorter than the time-scales for which the behaviour of the stochastic perturbations can be probed observationally, so that the assumed regularity of the velocity kicks or the specific value of the parameter $\Delta t$ have a negligible impact on our main results. 

{Let us now consider the standard driven damped harmonic oscillator, for which the driving acceleration is given by} 
\begin{equation}
a_f(t)=\tilde{a} \sin \left (\omega t + \varphi_{a_f}\right)\,,
\label{a_st}
\end{equation}
where $\tilde{a}$ is a constant amplitude of the acceleration associated to the driving mechanism of angular frequency $\omega$. {We shall refer to the model described by Eq.~(\ref{fho}) with $a_f(t)$ given by Eq.~(\ref{a_st}) as the externally driven damped harmonic oscillator. In section~\ref{modifieda}, a modification will be introduced to $a_f(t)$ which takes into account the explicit dependence on $x$ and ${\dot x}$ expected in the case of an internal forcing.}

In the absence of noise ($\xi=0$), after a transient phase dependent on initial conditions, the general solution of the driven damped harmonic oscillator tends to a steady state with
\begin{eqnarray}
x_f(t) &=& A_f  \sin \theta_f \,, \label{fhos1}\\ 
{\dot x_f(t)} &=& \omega A_f  \cos \theta_f \label{fhos2}\,, 
\end{eqnarray}
where 
\begin{eqnarray}
\theta_f&=&\omega t + \varphi_f\,, \\
A_f&=&\frac{\tilde a}{\omega \omega_*}\,, \\
\omega_* &=& \sqrt{ 4 \eta^2 + \frac{(\omega_0^2-\omega^2)^2}{\omega^2}}\,,\\
\varphi_*&=&\varphi_f - \varphi_{a_f} =\arctan \left(\frac{\omega_0^2 -\omega^2}{2 \omega \eta}\right)-\frac{\pi}{2}\,\label{phistar}\,,
\end{eqnarray}
and $\varphi_f$ is a constant.
The resonant angular frequency, defined as the value of $\omega$ for which the amplitude $A_f$ is maximal (or, equivalently, $\omega\omega_*$ is minimal) is 
\begin{equation}
\omega_r= \sqrt{\omega_0^2-2 \eta^2}\,.
\end{equation}

Throughout this work we shall assume that the driving frequency is equal to the resonant frequency ($\omega = \omega_r$), an assumption that ensures an optimal condition for the modes to be visible, for a given driving. In addition we shall consider that $\eta \ll \omega_0$ (implying that the characteristic damping time exceeds significantly the oscillation period, which is expected in general). In this case the driving frequency is close to the natural frequency ($\omega = \omega_r \approx \omega_0$) and $\varphi_* = -\pi/2$ to an excellent approximation. 

The solution to the equation of motion for the standard driven damped harmonic oscillator with white noise velocity perturbations may be written as $x=x_f + x_p$, where $x_f(t)$ is the steady state solution for the displacement for $\xi=0$ given in Eq. (\ref{fhos1}) and $x_p(t)$ is the perturbation induced by the noise. The evolution of $x_p$ is given by
\begin{equation}
{\ddot x}_p+2 \eta {\dot x}_p + \omega_0^2 x_p = \xi(t)\,.
\label{xp}
\end{equation}
In between random kicks $\xi=0$ and Eq.~(\ref{xp}) has the following general solution,
\begin{equation}
x_p(t) =  A_p(t) \sin\left( t\sqrt {\omega_0^2 - \eta^2} + \alpha \right) \,,
\end{equation}
 $A_p=Ce^{-\eta t}$, where $C$ is a constant, and $\alpha$ is an arbitrary phase.
Thus, in between kicks the full solution ($x(t)$, $\dot x(t)$) can still be written approximately as in Eqs~(\ref{fhos1})-(\ref{fhos2}), but with slowly-varying time-dependent amplitude, $A$, and phase, $\varphi$:
\begin{eqnarray}
x(t)&=& A(t)  \sin \left( \omega t + \varphi(t)\right)\,,  \label{fhos3} \\
{\dot x(t)} &=& \omega A(t)  \cos \left( \omega t  +\varphi(t)\right) \label{fhos4}\,.
\end{eqnarray}
In going from Eq.~(\ref{fhos3}) to Eq.~(\ref{fhos4}) we have taken into account the assumption previously discussed that $\eta \ll \omega_0 \approx \omega$, which implies that both $\dot A/A$ and $\dot\varphi$ are much smaller than $\omega$. 

The values of $A$ and $\varphi$ (or, equivalently, $A_p$ and $\alpha$) immediately after each kick may be determined from the corresponding values immediately before, knowing the values of $x$ and ${\dot x}$ prior to each kick as well as the velocity change $\Delta {\dot x}$ induced by each kick, and by requiring that $x$ is always continuous.
\begin{figure} 
	         \begin{minipage}{1.\linewidth}  
               \rotatebox{0}{\includegraphics[width=0.97\linewidth]{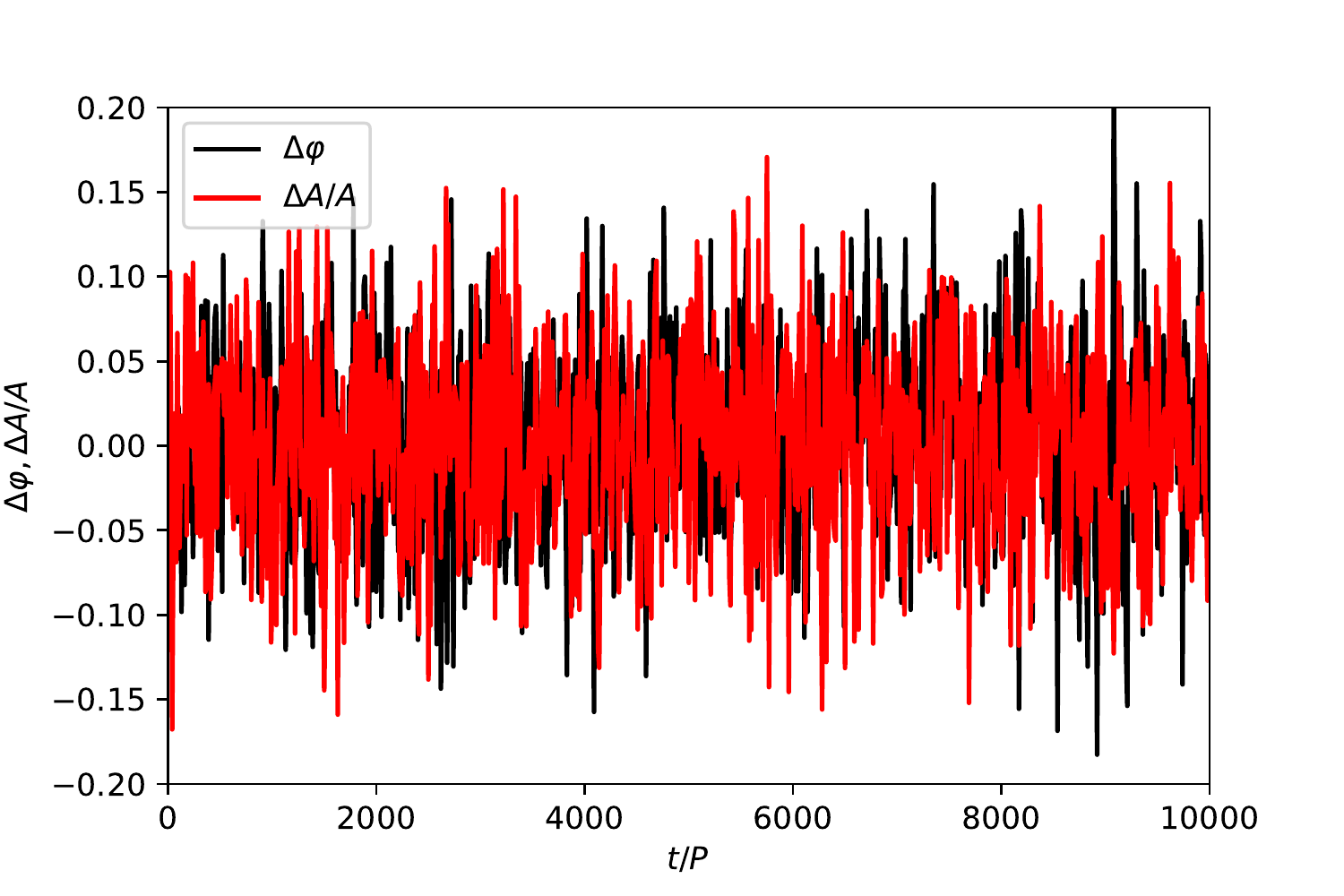}}
         \end{minipage}
	\caption{Phase variation $\Delta \varphi$ and relative amplitude variation $\Delta A/A$ (black and and red solid lines, respectively) as a function of the number of pulsation cycles $t/P$ for a single realization of the evolution of the externally driven damped harmonic oscillator with noise, assuming a total time span equal to  $10^4$ periods, $\eta = 0.2 P^{-1}$, and $A_N/A=0.01$.  {The variations are computed with respect to the constant phase and amplitude of the noiseless case (see text for details) and are shown at regular time intervals of $10$ periods.} The results show that, in the presence of a coherent external driving mechanism, a small amplitude stochastic excitation mechanism produces limited changes of similar magnitude to the phase $\varphi$ and amplitude $A$ of the oscillations.}
	\label{fig1}
\end{figure}
If $A_p \ll A_f$ then the solution for the displacement $x$ $(= x_f+x_p)$ is always very close to that of the standard driven damped harmonic oscillator without noise. In this case, the phase varies little, independently of the elapsed time. 

To illustrate this, we computed the solutions in the presence of noise. 
{The results are illustrated in Fig.~\ref{fig1}.} In particular, we show the evolution of the phase difference $\Delta\varphi\equiv\varphi-\varphi_f$ and of the relative amplitude variation $\Delta A/A\equiv(A-A_f)/A_f$ in terms of the number of elapsed pulsation cycles, taking $\Delta\varphi=0$ and $\Delta A=0$ at $t=0$. Both quantities vary little, as expected, because the phase $\varphi_{a_f}$ of the driving acceleration is fixed, as may happen in the case of an external forcing. However, the excitation mechanisms inside the stars are expected to depend explicitly on the phase $\varphi$ of the oscillation cycles, which, unlike in the case discussed in the present section, is, in general, time-dependent. In the following section we shall consider a more realistic assumption for the acceleration associated with the driving mechanism which is free from this problem.

\section{Internally Driven Damped Harmonic Oscillator with Noise: analytical results}
\label{modifieda}
As discussed in the previous section, in the presence of an internal driving mechanism the assumption that the acceleration associated to the driving mechanism in Eq. (\ref{fho}) always has the same phase in the presence of noise is unrealistic. Hence, we shall modify $a_f(t)$ in order to make it explicitly dependent on $x$ and $\dot x$ in such a way that, in the absence of noise, the steady state solution is unaltered. {We shall refer to this model, described by Eq.~(\ref{fho}) with the modified $a_f(t)$, as the internally driven damped harmonic oscillator}. Equations (\ref{fhos3}) and (\ref{fhos4}) imply that
\begin{equation}
\chi(x,\dot x) \equiv \frac{\dot x}{x \omega}= \cot \theta\,,
\end{equation}
where $\theta=\omega t+\varphi (t)$. 
Therefore, we shall now consider a forcing acceleration described by
\begin{equation}
a_f=\tilde{a}\sin\left(\theta(x,\dot x) -\varphi_* \right)\,,
\label{af_int}
\end{equation}
where $\varphi_*$ is given by Eq.~(\ref{phistar}), and
\begin{eqnarray}
\theta &=& {\rm arccot} \chi\,, \quad {\rm if} \  {\rm sign}(\dot x)=+1\,,\nonumber\\
\theta &=& \pi+ {\rm arccot} \chi\,, \quad {\rm if} \ {\rm sign}(\dot x)=-1\,.
\label{thechi}
\end{eqnarray}
Here, by convention, the range of the function ${\rm arccot}$ is $[0,\pi]$. When $\xi=0$, $\varphi(t)=\varphi_f= {\rm constant}$ and Eq.~(\ref{af_int}) reduces to Eq.~(\ref{a_st}). However, in the presence of noise, the forcing in each pulsation cycle now responds to the pulsation phase in that cycle, which may have been slightly modified from that in the previous cycle by the stochastic perturbations.  

On timescales much shorter than $\eta^{-1}$, the internally driven damped harmonic oscillator behaves essentially as a simple harmonic oscillator {($\eta=0$ and $a_f=0$)} with noise. Let us consider the impact in the phase and amplitude of the oscillations of a single instantaneous random kick $\Delta {\dot x}$, given at the time $t_k$, in the velocity of the oscillations ($\dot x \to \dot x + \Delta {\dot x}$). The displacement and the velocity of the harmonic oscillator is given approximately by Eqs. (\ref{fhos3}) and Eqs. (\ref{fhos4}), with $A=A_-$ and $\theta=\theta_-=\omega t + \varphi_-$, for $t<t_k$, and $A=A_+$ and $\theta_+=\omega t + \varphi_+$, for $t>t_k$. Here, we shall again assume that the variation of the amplitude with respect to the amplitude $A_f$ of the steady state solution for the forced harmonic oscillator in the absence of noise is small.
Therefore, 
\begin{eqnarray}
\Delta \chi (t_k)= \frac{\dot x (t_{k+}) - \dot x (t_{k-})}{\omega x (t_{k})} &=& \frac{\Delta \dot x (t_k)}{\omega x [t_{k}]}= \cot \theta_+ - \cot \theta_- \nonumber \\ 
&\approx& -\frac{\Delta \varphi (t_k)}{\sin^2 \theta_-}\,,
\end{eqnarray}
where $\Delta \varphi (t_k) ={\theta}_+-\theta_-$ is the change in the phase of the oscillation associated to the instantaneous random kick at time $t_k$, and the approximation is valid up to first order in $\Delta \varphi (t_k)$. Hence, using Eq. (\ref{fhos3}) with $A=A_-$ and $\theta=\theta_-$, the change in the phase at $t=t_k$ is given by
\begin{equation}
\Delta \varphi (t_k) \approx - \frac{\Delta {\dot x} \sin \theta_-}{A \omega}\,. \label{dph}
\end{equation}

The change in amplitude of the oscillations may be calculated by requiring that $x(t)$ is always a continuous function. The requirement of continuity at $t=t_i$ implies that
\begin{equation}
A_-   \sin \theta_- = A_+ \sin \theta_+\,,
\end{equation}
or equivalently that
\begin{eqnarray}
(A_+ - A_-) \sin \theta_-&=& - A_{+}  (\sin \theta_+ -  \sin \theta_-) \nonumber \\ 
&\approx& -A_+ \Delta \varphi (t_k) \cos \theta_-\,,
\end{eqnarray}
where the approximation is valid up to first order in $\Delta \varphi (t_k)$. Therefore, 
\begin{equation}
\frac{\Delta A}{A}(t_k) \approx - \Delta \varphi (t_k) \cot \theta_-\,, \label{da}
\end{equation}
up to first order in $\Delta A/A (t_k)$ and $\Delta \varphi (t_k)$, {where $\Delta A(t_k)=A_+-A_-$}.

The root mean square variation of the phase due to a single random kick at an arbitrary time $t_k$ is given by
\begin{eqnarray}
\sigma_{\Delta \varphi(t_k)} &\equiv&  \sqrt{\left \langle \left(\Delta \varphi (t_k)\right)^2 \right \rangle}  \nonumber \\ 
&=& \sigma_{\Delta A(t_k)/A} \equiv \sqrt{ \left \langle \left(\frac{\Delta A}{A} (t_k) \right)^2 \right \rangle} \nonumber \\
&\approx& \sqrt{\frac{\left \langle \left(\Delta {\dot x}(t_k) \right)^2 \right \rangle \left \langle \left(\sin\theta_- \right)^2 \right \rangle}{\omega^2 A^2}} \nonumber \\
&\approx& \frac{A_N \sqrt{2\omega \Delta t}}{A}\,,\label{sigma}
\end{eqnarray}
where we have taken into account that $\left \langle \left(\sin \theta_- \right)^2 \right \rangle =1/2$ and assumed that $\Delta {\dot x} (t_k) = 2 r  A_{N} \omega_0 \sqrt {\omega_0 \Delta t}$ with $\omega_0\approx \omega$, and, in a accordance to the discussion in section~\ref{standard}, $r$ is a random variable with a normal distribution of mean zero and unit standard deviation ($r \sim N(0,1)$).

In a total observing time $t_{\rm obs}$, the average number of kicks is equal to $N=t_{\rm obs}/\Delta t$, each described by Eq.~(\ref{sigma}). The phase variation displays a random walk and, therefore, the root mean square variance of the phase at the end of that time is given by
\begin{equation}
\sigma_{\Delta \varphi} = \sqrt{N} \sigma_{\Delta \varphi(t_k)} \approx \frac{A_N}{A} \sqrt{2\omega t_{\rm obs}} \,.
\label{sigphi}
\end{equation}
 The random walk nature of the phase variation $\Delta \varphi$ gives rise to a red noise power spectra of $\Delta \varphi$ which we shall discuss in the following section.

In order to estimate the root mean square of $\Delta A/A$ in the case of periodic instantaneous random kicks, we take into account that the amplitude perturbation generated at an arbitrary time $t_k$ decays roughly proportionally to $e^{-\eta t}$, meaning that the memory of previous amplitude perturbations is effectively erased on a timescale of the order of $\eta^{-1}$. Therefore
\begin{eqnarray}
\sigma^2_{\Delta A/A} &\approx& \frac{1}{\Delta t}\int_0^{t_{\rm obs}} \sigma^2_{\Delta A/A(t_k)} e^{-2 \eta t} dt \nonumber \\
&\approx&  \left(\frac{A_N}{A}\right)^2 \frac{\omega}{\eta}\left(1-e^{-2 \eta t_{\rm obs}} \right) \,.
\label{siga}
\end{eqnarray}
For $t_{\rm obs} \gg \eta^{-1}$ the root-mean-square of $\Delta A/A$ is approximately given by
\begin{equation}
\sigma_{\Delta A/A} =  \frac{A_N}{A} \sqrt{\frac{\omega}{\eta}}\,.
\label{siga_lim}
\end{equation}
{From Eqs~(\ref{sigphi})--(\ref{siga_lim}), we then find that, $\sigma_{\Delta \varphi} \approx \sigma_{\Delta A/A}$ 
for $t_{\rm obs} \lesssim \eta^{-1}$,
and $\sigma_{\Delta \varphi} \approx \sigma_{\Delta A/A} (2t_{\rm obs} \eta)^{1/2}$
for $t_{\rm obs} \gg \eta^{-1}$. {These analytical results are valid under the assumption that the relative amplitude variations are small}.}

\begin{figure} 
	         \begin{minipage}{1.\linewidth}  
               \rotatebox{0}{\includegraphics[width=0.97\linewidth]{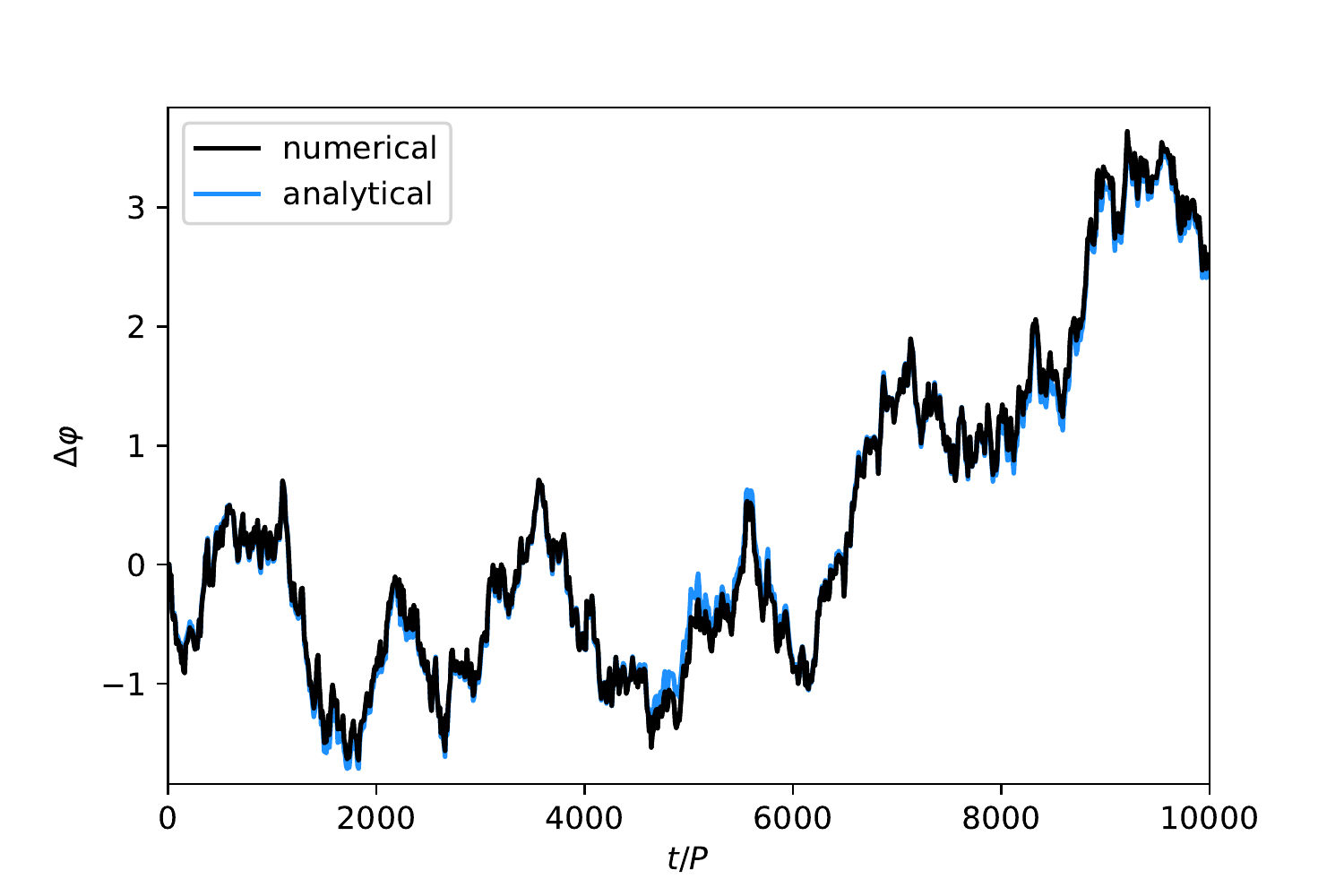}}
         \end{minipage}
	\caption{Phase variation $\Delta\varphi$ as a function of the number of oscillation cycles $t/P$ for a single realization of the evolution of the internally driven damped harmonic oscillator with noise, assuming a total time span equal to  $10^4$ periods, $\eta = 0.2 P^{-1}$, and $A_N/A=0.01$ (the numerical and analytical results are given by the black and blue solid lines, respectively). The variation is shown at regular time intervals of $10$ periods. The results show that, in the presence of a coherent internal driving mechanism, a small amplitude stochastic excitation mechanism is able to produce, over a sufficient amount time, large variations in the phase of the oscillations.}
	\label{fig2}
\end{figure}

\section{Internally Driven Damped Harmonic Oscillator with Noise: numerical results}
\label{modifiedn}
In order to investigate the impact of stochastic perturbations on classical pulsators, we computed the evolution of the displacement $x$ by numerically solving the equation of motion for the internally driven damped harmonic oscillator with white noise velocity perturbations using a fourth order Runge-Kutta algorithm. The initial conditions are those of the steady solution of the standard driven damped harmonic oscillator (with $\xi=0$).

Figure ~\ref{fig2} illustrates the phase variation $\Delta\varphi$ as a function of the number of oscillation cycles $t/P$ (the values of $\Delta \varphi$ are displayed with a cadence of $10$ periods of oscillation) for a single numerical realization of the evolution of an internally driven damped harmonic oscillator with noise, considering a total time span equal to  $10^4$ periods, $\eta = 0.2 P^{-1}$, and $A_N/A=0.01$ (the time interval between successive kicks was assumed to be equal to $\Delta t = 0.025P$). Note that $\omega_0 \sim 2 \pi P^{-1}$, thus implying that $\eta / \omega_0 \sim 0.03$ is significantly smaller than unity for the value of $\eta$ adopted in this realization. The evolution of the phase variation $\Delta \varphi$ obtained by solving numerically Eq. (\ref{fho}) is shown in black, while that resulting from the analytical approximation given in Eq. (\ref{dph}) for the instantaneous phase variation associated to each kick is shown in blue. Here, $\Delta \varphi$ was assumed to vanish at $t=0$. Figure ~\ref{fig2} shows that, not only Eq. (\ref{dph}) provides an excellent approximation which may be used to accurately compute the phase variation $\Delta \varphi$ over many oscillation cycles, but also that in the presence of a coherent internal driving mechanism, small amplitude stochastic perturbations are able to produce, over a sufficient amount of time, large variations in the phase of the oscillations. This is in contrast with the results for the phase variation $\Delta \varphi$ of a {damped harmonic oscillator with noise driven by an external forcing shown if Fig. ~\ref{fig1}}.

\begin{figure} 
	         \begin{minipage}{1.\linewidth}  
               \rotatebox{0}{\includegraphics[width=0.97\linewidth]{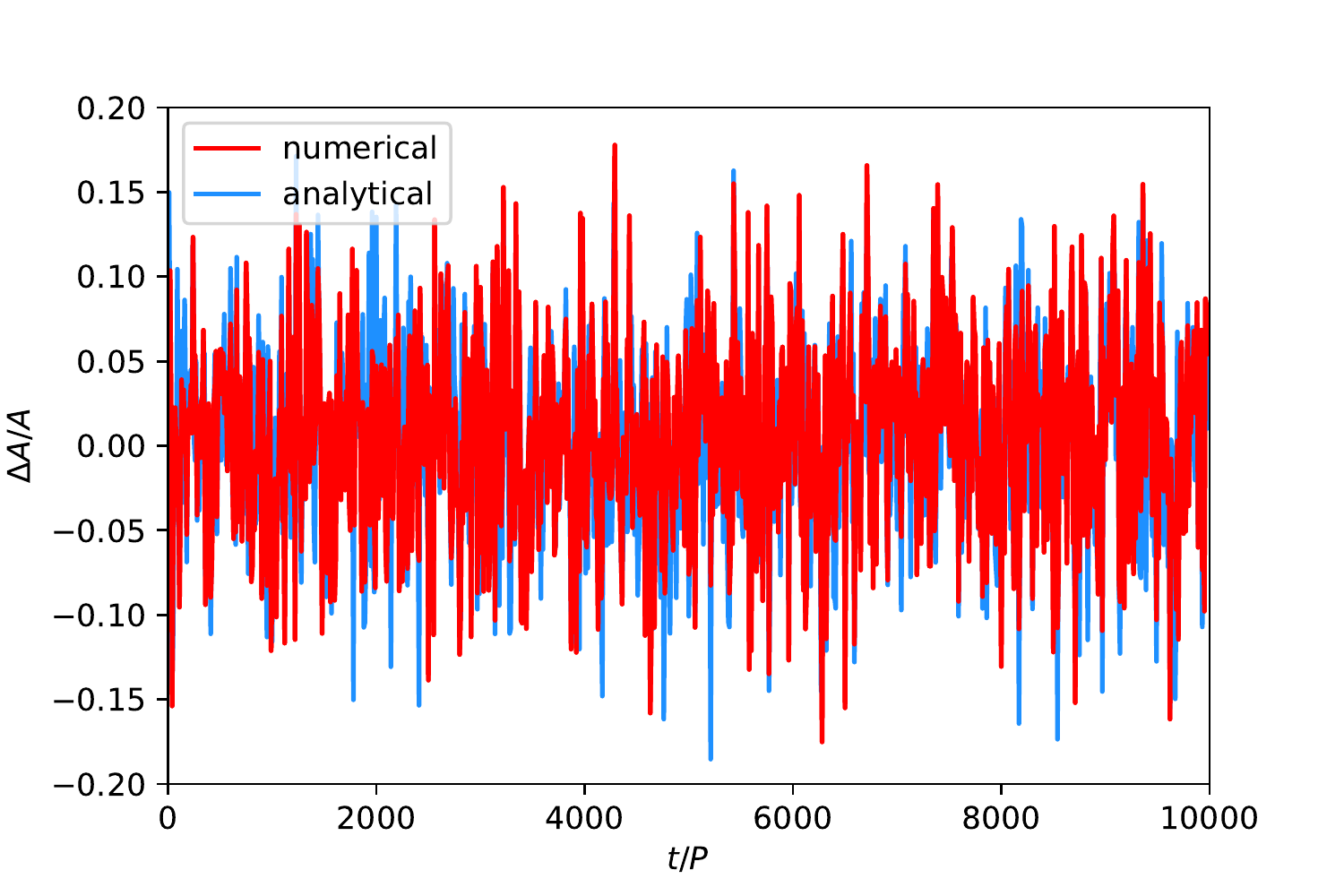}}
         \end{minipage}
	\caption{Relative amplitude variation $\Delta A/A$ as a function of the number of oscillation cycles $t/P$ for the same realization of the evolution of the internally driven damped harmonic oscillator with noise considered in Fig. ~\ref{fig2} (the numerical and analytical results are given by the red and blue solid lines, respectively). The variation is shown at regular time intervals of $10$ periods. In the case of $\Delta A/A$ the results are similar to those obtained for the externally driven damped harmonic oscillator with noise.}
	\label{fig3}
\end{figure}

Figure ~\ref{fig3} illustrates the relative amplitude variation $\Delta A/A$ as a function of the number of oscillation cycles $t/P$ for the same realization of the evolution of the internally driven damped harmonic oscillator with noise considered in Fig. ~\ref{fig2}. The evolution of $\Delta A/A$ obtained by solving numerically Eq. (\ref{fho}) and then determining the amplitude from the maxima of $x$ over timescales of the order of the oscillation period $P$ is shown in red, while that obtained by using the approximation given in Eq. (\ref{da}) for the instantaneous amplitude variation associated to each kick in combination with a subsequent exponential decay proportional to $e^{-\eta t}$ is shown in blue. Figure ~\ref{fig3} shows that both these results for $\Delta A/A$ are similar to the ones obtained in the previous section for the externally driven damped harmonic oscillator with white noise velocity perturbations.

\begin{figure} 
	         \begin{minipage}{1.\linewidth}  
               \rotatebox{0}{\includegraphics[width=0.97\linewidth]{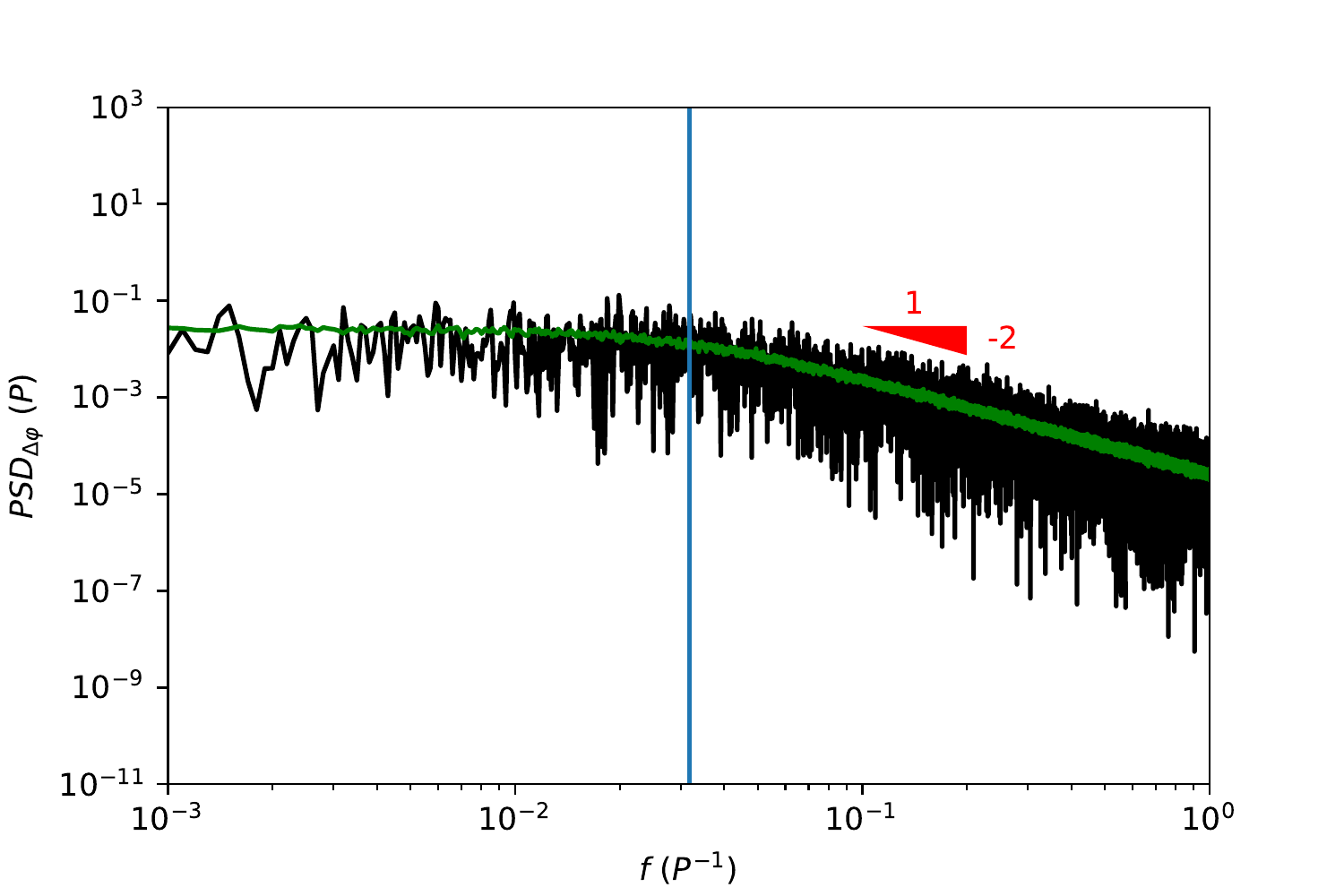}}
               \rotatebox{0}{\includegraphics[width=0.97\linewidth]{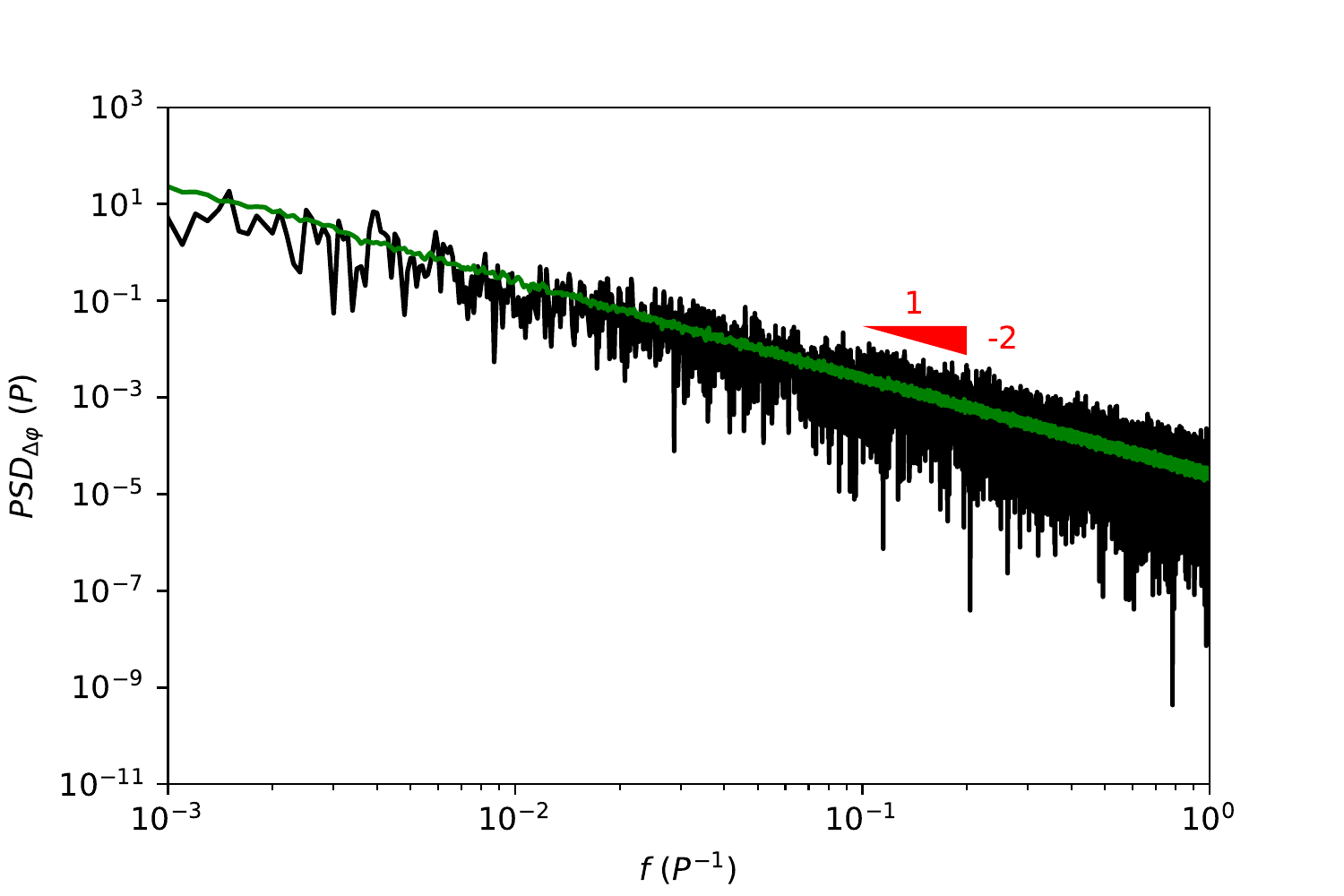}}
         \end{minipage}
	\caption{Power spectral density of the phase variation $\Delta \varphi$ obtained considering the externally (upper panel) and the internally (lower panel) driven damped harmonic oscillator with noise. The black line represents the results obtained for a single realization (the same that was considered in Figs. ~\ref{fig2} and ~\ref{fig3}) while the green line represents the average over $100$ realizations. The hypotenuse of the red triangle has a $f^{-2}$ slope characteristic of red noise. Notice that the transition from red-noise to white-noise (flat-spectrum) at $f \sim \eta /(2 \pi)$ (the blue vertical line in the upper panel is defined by $f=\eta/(2\pi)$), does not happens in the modified case. }
	\label{fig4}
\end{figure}

{The different evolution of the phase variation $\Delta \varphi$ in the context of standard (external) and modified (internal) versions of the driven damped harmonic oscillator with white noise velocity perturbations is also imprinted in the power spectrum of the phase variation $\Delta \varphi$.} In the internally driven case $\Delta \varphi$ displays a random walk over arbitrary large time spans, which leaves a red noise signature in the power spectrum. On the other hand, in the externally driven case the random walk of $\Delta \varphi$ is only approximately valid on time scales much smaller than $\eta^{-1}$, with results separated by a larger time difference being essentially independent. Hence, in the case of the externally driven damped harmonic oscillator with white noise velocity perturbations a transition from a large frequency red noise spectrum to a low frequency white noise spectrum at a frequency approximately equal to $2\pi/\eta$ would be expected. Figure ~\ref{fig4} shows that this is indeed the case. It displays the power spectral density of the phase variation $\Delta \varphi$ obtained for the evolution of the externally driven and internally driven damped harmonic oscillators with white noise velocity perturbations (upper and lower panels, respectively). The black line represents the results obtained for a single realization (the same realization considered in Figs. ~\ref{fig2} and ~\ref{fig3}) while the green line represents the average over $100$ realizations. The hypotenuse of the red triangle has a $f^{-2}$ slope characteristic of red noise. Notice that the transition from red noise to white noise (flat spectrum) at $f \sim \eta/(2\pi)$ (indicated by the blue vertical line in the upper panel of Fig. ~\ref{fig4}), does not happen in the internally driven case.

\begin{figure} 
	         \begin{minipage}{1.\linewidth}  
               \rotatebox{0}{\includegraphics[width=0.97\linewidth]{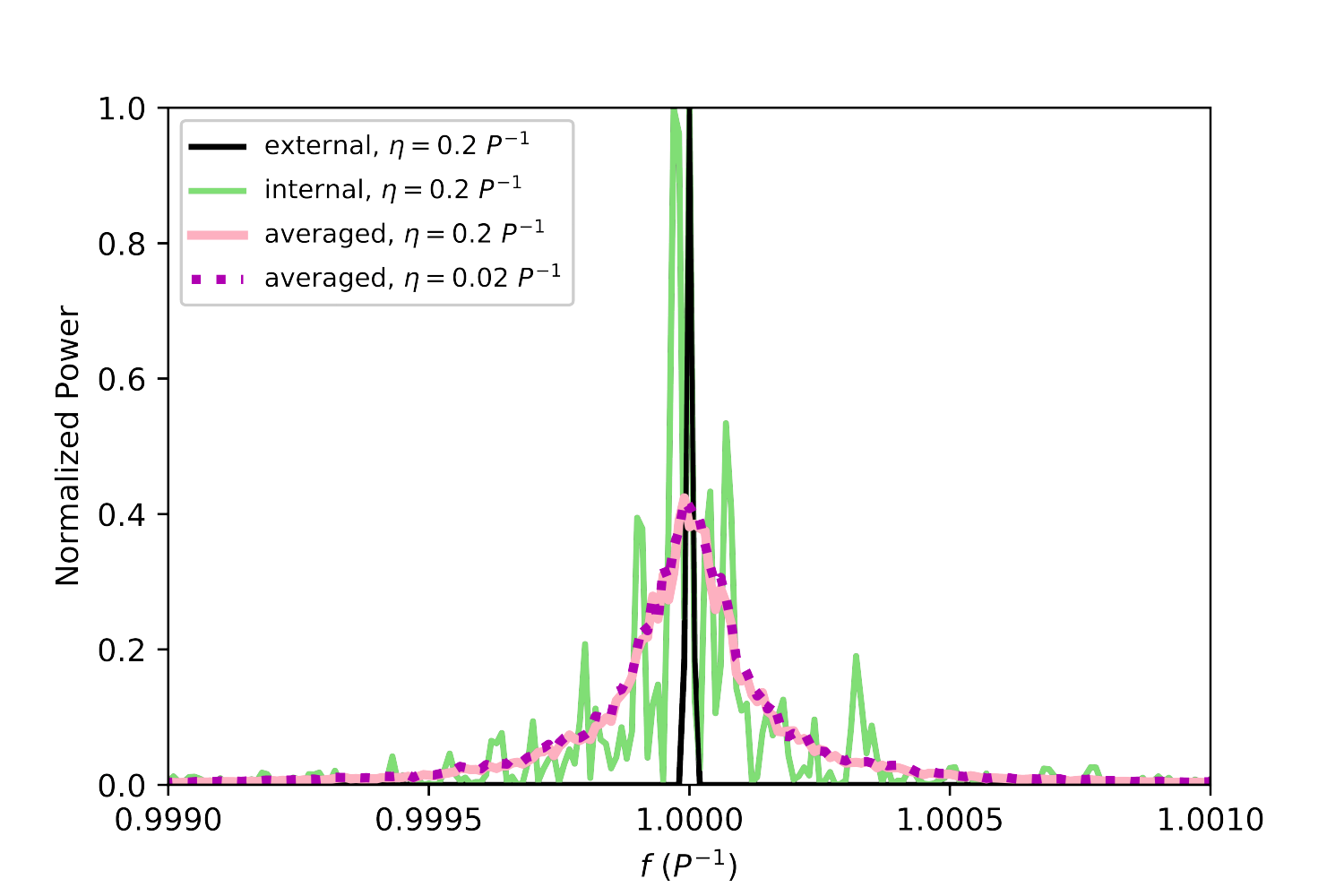}}
         \end{minipage}
	\caption{{Power spectra of the velocity for a single numerically realization assuming a time span equal to $10^5$ periods, $\eta = 0.2 P^{-1}$, and $A_N/A=0.01$, considering the externally (black line) and the internally (light green line) driven damped harmonic oscillator with noise. Both power spectra are normalized such that their maximum amplitude is equal to one. The ragged erratic appearance displayed by the power spectrum generated assuming an internal forcing mechanism contrasts with the sharp high amplitude peak obtained for the externally driven damped harmonic oscillator. The continuous pink and dashed magenta smooth lines show the results from averaging over 100 realizations of the internally driven damped harmonic oscillator for the same model parameters and when taking $\eta=0.02P^{-1}$, respectively. The independence of the average curve on $\eta$ in the regime of small amplitude perturbations considered here is evident from the overlap of the two lines.}}
	\label{fig5}
\end{figure}

Figure ~\ref{fig5} shows the power spectra of the velocity for a single numerically realization of a time series with $10^5$ oscillation cycles, $\eta = 0.2 P^{-1}$, and $A_N/A=0.01$, considering the externally and the internally driven damped harmonic oscillator with noise (black and light green lines, respectively). Here, for vizualization purposes, both power spectra are normalized in such a way that their maximum amplitude is equal to unity. Notice that the ragged erratic appearance displayed by the power spectrum generated assuming an internal forcing mechanism contrasts with the sharp high amplitude peak obtained for the externally driven damped harmonic oscillator. {Also shown in the figure is the average performed over 100 realizations for the case of the internally driven damped harmonic oscillator with the same parameters as above (smooth continuous pink line). For comparison, we present also the average of another set of 100 realizations for the same model computed with a damping constant 10 times smaller ($\eta=0.02P^{-1}$; smooth dashed magenta line). The two lines clearly overlap, confirming that, as long as the relative amplitude variations are small, the envelope of the power is essentially independent of $\eta$. }

\section{Conclusions}
\label{conclusions}

We developed a parametric phenomenological model to describe the impact of stochastic perturbations on classical pulsators based upon a driven damped harmonic oscillator with added white noise. {Besides $\omega_0$, which in our model is approximately equal to the pulsation frequency, hence directly determined by the observations, the model incorporates three parameters, $A_N/A$, characterising the relative amplitude of the white noise, $\eta$, the inverse of the damping time, and $\Delta t$, the interval between consecutive stochastic kicks. However, the choice of $\Delta t$ is unimportant, so far as it is significantly smaller than the time scale over which the stochastic signature can be observationally probed. Thus the results can effectively be thought of as pertaining to a two-parameter model.}

{We started by considering the standard case of an external coherent (resonant) forcing, which would be adequate if the driving phase was unaltered by the dynamics of the stellar interior (as expected, {\it e.g.}, in tidally-excited oscillations \citep{zahn70}), and then modified it by replacing the forcing term by an internal (resonant) forcing which reduces to the standard one in the absence of noise. Thus, the modified model considers that the driving throughout any given cycle is adjusted to the current phase of the oscillation which may differ from the phase of previous cycles due to small perturbations introduced by the stochastic kicks.}

{The model predictions for the evolution of the phase and relative amplitude variations ($\Delta \varphi$ and ${\Delta A/A}$, respectively) have been studied using both analytical approximations and numerical simulations exploring, in particular, the case in which the relative amplitude variations remain small. We have shown that in the case of an internal forcing, {the rms value of the phase variation $\Delta \varphi$ induced by the stochastic perturbations grows on average proportionally to the square root of the observing time.} This is in contrast with the relative amplitude variations whose rms stabilises after a time $\gtrsim \eta^{-1}$, the two observables being related by $\sigma_{\Delta \varphi} = \sigma_{\Delta A/A} (2t_{\rm obs} \eta)^{1/2}$, for $t_{\rm obs} \gg \eta^{-1}$.}

{We have further shown that in the case of an internal forcing the signatures of the random walk evolution of the phase variation $\Delta \varphi$ are imprinted on its characteristic ragged erratic red noise power spectrum --- the power spectrum of the signal (${\dot x}$)  also displaying a ragged erratic appearance. An important outcome of the model, is that, for a given observing time, the rms of the phase depends only on one model parameter, $A_N/A$, opening the interesting possibility of that parameter being directly constrained from either an O-C diagram \citep{sterken05} or the power spectrum of the signal.} {A related fact, also worth emphasising, is that the envelope of the power spectrum of the signal in this case is independent of the damping time, as shown in Fig.~\ref{fig5}. This is valid as long as the relative amplitude variations remain small and is in clear contrast with what is observed is solar-like pulsators, where the damping time can be inferred from the width of the Lorentzian profile fitted to the power spectrum.}

{Our results imply that, given sufficient time, the variations of the phase associated to the stochastic perturbations will always become sufficiently large to be probed observationally in internally excited classical pulsators.}

\section*{Acknowledgements}
P. P. A. thanks the support from FCT -- Funda{\c c}\~ao para a Ci\^encia e a Tecnologia -- through the Sabbatical Grant No. SFRH/BSAB/150322/2019. M. S. Cunha is supported by national funds through FCT in the form of a work contract. This work was supported by FCT through national funds (PIDDAC) (grants: PTDC/FIS-AST/30389/2017 and UID/FIS/04434/2019) and by FEDER - Fundo Europeu de Desenvolvimento Regional through COMPETE2020 - Programa Operacional Competitividade e Internacionalização (grant: POCI-01-0145-FEDER-030389). W.~J.~C. acknowledges support from the UK Science and Technology Facilities Council (STFC). Funding for the Stellar Astrophysics Centre is provided by The Danish National Research Foundation (Grant agreement no.: DNRF106).




\bibliography{solar-like.bib} 

\begin{thebibliography}{}
\makeatletter
\relax
\def\mn@urlcharsother{\let\do\@makeother \do\$\do\&\do\#\do\^\do\_\do\%\do\~}
\def\mn@doi{\begingroup\mn@urlcharsother \@ifnextchar [ {\mn@doi@}
  {\mn@doi@[]}}
\def\mn@doi@[#1]#2{\def\@tempa{#1}\ifx\@tempa\@empty \href
  {http://dx.doi.org/#2} {doi:#2}\else \href {http://dx.doi.org/#2} {#1}\fi
  \endgroup}
\def\mn@eprint#1#2{\mn@eprint@#1:#2::\@nil}
\def\mn@eprint@arXiv#1{\href {http://arxiv.org/abs/#1} {{\tt arXiv:#1}}}
\def\mn@eprint@dblp#1{\href {http://dblp.uni-trier.de/rec/bibtex/#1.xml}
  {dblp:#1}}
\def\mn@eprint@#1:#2:#3:#4\@nil{\def\@tempa {#1}\def\@tempb {#2}\def\@tempc
  {#3}\ifx \@tempc \@empty \let \@tempc \@tempb \let \@tempb \@tempa \fi \ifx
  \@tempb \@empty \def\@tempb {arXiv}\fi \@ifundefined
  {mn@eprint@\@tempb}{\@tempb:\@tempc}{\expandafter \expandafter \csname
  mn@eprint@\@tempb\endcsname \expandafter{\@tempc}}}

\bibitem[\protect\citeauthoryear{{Aerts}, {Christensen-Dalsgaard}  \&
  {Kurtz}}{{Aerts} et~al.}{2010}]{aerts10}
{Aerts} C.,  {Christensen-Dalsgaard} J.,   {Kurtz} D.~W.,  2010,
  {Asteroseismology}

\bibitem[\protect\citeauthoryear{{Balona}, {Holdsworth}  \& {Cunha}}{{Balona}
  et~al.}{2019}]{balona2019}
{Balona} L.~A.,  {Holdsworth} D.~L.,   {Cunha} M.~S.,  2019, \mn@doi [\mnras]
  {10.1093/mnras/stz1423}, \href
  {https://ui.adsabs.harvard.edu/abs/2019MNRAS.487.2117B} {487, 2117}

\bibitem[\protect\citeauthoryear{{Bedding}, {Kiss}, {Kjeldsen}, {Brewer},
  {Dind}, {Kawaler}  \& {Zijlstra}}{{Bedding} et~al.}{2005}]{bedding05}
{Bedding} T.~R.,  {Kiss} L.~L.,  {Kjeldsen} H.,  {Brewer} B.~J.,  {Dind} Z.~E.,
   {Kawaler} S.~D.,   {Zijlstra} A.~A.,  2005, \mn@doi [\mnras]
  {10.1111/j.1365-2966.2005.09281.x}, \href
  {https://ui.adsabs.harvard.edu/abs/2005MNRAS.361.1375B} {361, 1375}

\bibitem[\protect\citeauthoryear{{Benk{\H{o}}}, {Jurcsik}  \&
  {Derekas}}{{Benk{\H{o}}} et~al.}{2019}]{benko19}
{Benk{\H{o}}} J.~M.,  {Jurcsik} J.,   {Derekas} A.,  2019, \mn@doi [\mnras]
  {10.1093/mnras/stz833}, \href
  {https://ui.adsabs.harvard.edu/abs/2019MNRAS.485.5897B} {485, 5897}

\bibitem[\protect\citeauthoryear{{Borucki} et~al.,}{{Borucki}
  et~al.}{2010}]{borucki10}
{Borucki} W.~J.,  et~al., 2010, \mn@doi [Science] {10.1126/science.1185402},
  \href {https://ui.adsabs.harvard.edu/abs/2010Sci...327..977B} {327, 977}

\bibitem[\protect\citeauthoryear{{Bowman}, {Kurtz}, {Breger}, {Murphy}  \&
  {Holdsworth}}{{Bowman} et~al.}{2016}]{bowman2016}
{Bowman} D.~M.,  {Kurtz} D.~W.,  {Breger} M.,  {Murphy} S.~J.,   {Holdsworth}
  D.~L.,  2016, \mn@doi [\mnras] {10.1093/mnras/stw1153}, \href
  {https://ui.adsabs.harvard.edu/abs/2016MNRAS.460.1970B} {460, 1970}

\bibitem[\protect\citeauthoryear{{Breger} \& {Bischof}}{{Breger} \&
  {Bischof}}{2002}]{breger02}
{Breger} M.,  {Bischof} K.~M.,  2002, \mn@doi [\aap]
  {10.1051/0004-6361:20020124}, \href
  {https://ui.adsabs.harvard.edu/abs/2002A&A...385..537B} {385, 537}

\bibitem[\protect\citeauthoryear{{Breger} \& {Pamyatnykh}}{{Breger} \&
  {Pamyatnykh}}{1998}]{breger98}
{Breger} M.,  {Pamyatnykh} A.~A.,  1998, \aap, \href
  {https://ui.adsabs.harvard.edu/abs/1998A&A...332..958B} {332, 958}

\bibitem[\protect\citeauthoryear{{Breger}, {Montgomery}, {Lenz}  \&
  {Pamyatnykh}}{{Breger} et~al.}{2017}]{breger2017}
{Breger} M.,  {Montgomery} M.~H.,  {Lenz} P.,   {Pamyatnykh} A.~A.,  2017,
  \mn@doi [\aap] {10.1051/0004-6361/201629797}, \href
  {https://ui.adsabs.harvard.edu/abs/2017A&A...599A.116B} {599, A116}

\bibitem[\protect\citeauthoryear{{Buchler}, {Goupil}  \& {Hansen}}{{Buchler}
  et~al.}{1997}]{buchler1997}
{Buchler} J.~R.,  {Goupil} M.~J.,   {Hansen} C.~J.,  1997, \aap, \href
  {https://ui.adsabs.harvard.edu/abs/1997A&A...321..159B} {321, 159}

\bibitem[\protect\citeauthoryear{{Buchler}, {Koll{\'a}th}  \&
  {Cadmus}}{{Buchler} et~al.}{2004}]{buchler04}
{Buchler} J.~R.,  {Koll{\'a}th} Z.,   {Cadmus} Robert~R. J.,  2004, \mn@doi
  [\apj] {10.1086/422903}, \href
  {https://ui.adsabs.harvard.edu/abs/2004ApJ...613..532B} {613, 532}

\bibitem[\protect\citeauthoryear{{Chaplin} \& {Miglio}}{{Chaplin} \&
  {Miglio}}{2013}]{chaplin13}
{Chaplin} W.~J.,  {Miglio} A.,  2013, \mn@doi [\araa]
  {10.1146/annurev-astro-082812-140938}, \href
  {http://adsabs.harvard.edu/abs/2013ARA%26A..51..353C} {51, 353}

\bibitem[\protect\citeauthoryear{{Christensen-Dalsgaard}, {Kjeldsen}  \&
  {Mattei}}{{Christensen-Dalsgaard} et~al.}{2001}]{JCD01}
{Christensen-Dalsgaard} J.,  {Kjeldsen} H.,   {Mattei} J.~A.,  2001, \mn@doi
  [\apjl] {10.1086/338194}, \href
  {https://ui.adsabs.harvard.edu/abs/2001ApJ...562L.141C} {562, L141}

\bibitem[\protect\citeauthoryear{{Cunha} et~al.,}{{Cunha}
  et~al.}{2007}]{cunhaetal07}
{Cunha} M.~S.,  et~al., 2007, \mn@doi [Astronomy and Astrophysics Review]
  {10.1007/s00159-007-0007-0}, \href
  {http://adsabs.harvard.edu/abs/2007A%26ARv..14..217C} {14, 217}

\bibitem[\protect\citeauthoryear{{Degroote} et~al.,}{{Degroote}
  et~al.}{2010}]{degroote10}
{Degroote} P.,  et~al., 2010, \mn@doi [\nat] {10.1038/nature08864}, \href
  {http://adsabs.harvard.edu/abs/2010Natur.464..259D} {464, 259}

\bibitem[\protect\citeauthoryear{{Derekas} et~al.,}{{Derekas}
  et~al.}{2017}]{derekas17}
{Derekas} A.,  et~al., 2017, \mn@doi [\mnras] {10.1093/mnras/stw2399}, \href
  {https://ui.adsabs.harvard.edu/abs/2017MNRAS.464.1553D} {464, 1553}

\bibitem[\protect\citeauthoryear{{Eddington} \& {Plakidis}}{{Eddington} \&
  {Plakidis}}{1929}]{eddington1929}
{Eddington} A.~S.,  {Plakidis} S.,  1929, \mn@doi [\mnras]
  {10.1093/mnras/90.1.65}, \href
  {https://ui.adsabs.harvard.edu/abs/1929MNRAS..90...65E} {90, 65}

\bibitem[\protect\citeauthoryear{{Freytag}, {Liljegren}  \&
  {H{\"o}fner}}{{Freytag} et~al.}{2017}]{freytag17}
{Freytag} B.,  {Liljegren} S.,   {H{\"o}fner} S.,  2017, \mn@doi [\aap]
  {10.1051/0004-6361/201629594}, \href
  {https://ui.adsabs.harvard.edu/abs/2017A&A...600A.137F} {600, A137}

\bibitem[\protect\citeauthoryear{{Gilliland} et~al.,}{{Gilliland}
  et~al.}{2010}]{gillilandetal10}
{Gilliland} R.~L.,  et~al., 2010, \mn@doi [\pasp] {10.1086/650399}, \href
  {http://adsabs.harvard.edu/abs/2010PASP..122..131G} {122, 131}

\bibitem[\protect\citeauthoryear{{Holdsworth}, {Smalley}, {Kurtz},
  {Southworth}, {Cunha}  \& {Clubb}}{{Holdsworth} et~al.}{2014}]{holdworth14}
{Holdsworth} D.~L.,  {Smalley} B.,  {Kurtz} D.~W.,  {Southworth} J.,  {Cunha}
  M.~S.,   {Clubb} K.~I.,  2014, \mn@doi [\mnras] {10.1093/mnras/stu1303},
  \href {https://ui.adsabs.harvard.edu/abs/2014MNRAS.443.2049H} {443, 2049}

\bibitem[\protect\citeauthoryear{{Kilkenny}}{{Kilkenny}}{2010}]{kilkenny10}
{Kilkenny} D.,  2010, \mn@doi [\apss] {10.1007/s10509-010-0324-z}, \href
  {https://ui.adsabs.harvard.edu/abs/2010Ap&SS.329..175K} {329, 175}

\bibitem[\protect\citeauthoryear{{Kurtz}, {van Wyk}, {Roberts}, {Marang},
  {Handler}, {Medupe}  \& {Kilkenny}}{{Kurtz} et~al.}{1997}]{kurtz97}
{Kurtz} D.~W.,  {van Wyk} F.,  {Roberts} G.,  {Marang} F.,  {Handler} G.,
  {Medupe} R.,   {Kilkenny} D.,  1997, \mn@doi [\mnras]
  {10.1093/mnras/287.1.69}, \href
  {https://ui.adsabs.harvard.edu/abs/1997MNRAS.287...69K} {287, 69}

\bibitem[\protect\citeauthoryear{{Neilson}, {Percy}  \& {Smith}}{{Neilson}
  et~al.}{2016}]{neilson2016}
{Neilson} H.~R.,  {Percy} J.~R.,   {Smith} H.~A.,  2016, Journal of the
  American Association of Variable Star Observers (JAAVSO), \href
  {https://ui.adsabs.harvard.edu/abs/2016JAVSO..44..179N} {44, 179}

\bibitem[\protect\citeauthoryear{{Percy} \& {Colivas}}{{Percy} \&
  {Colivas}}{1999}]{percy1999}
{Percy} J.~R.,  {Colivas} T.,  1999, \mn@doi [\pasp] {10.1086/316290}, \href
  {https://ui.adsabs.harvard.edu/abs/1999PASP..111...94P} {111, 94}

\bibitem[\protect\citeauthoryear{{Pigulski} \& {Pojma{\'n}ski}}{{Pigulski} \&
  {Pojma{\'n}ski}}{2008}]{pigulski08}
{Pigulski} A.,  {Pojma{\'n}ski} G.,  2008, \mn@doi [\aap]
  {10.1051/0004-6361:20078580}, \href
  {https://ui.adsabs.harvard.edu/abs/2008A&A...477..907P} {477, 907}

\bibitem[\protect\citeauthoryear{{Smolec}}{{Smolec}}{2017}]{smolec17}
{Smolec} R.,  2017, \mn@doi [\mnras] {10.1093/mnras/stx679}, \href
  {https://ui.adsabs.harvard.edu/abs/2017MNRAS.468.4299S} {468, 4299}

\bibitem[\protect\citeauthoryear{{Sterken}}{{Sterken}}{2005}]{sterken05}
{Sterken} C.,  2005, {The O-C Diagram: Basic Procedures}.
p.~3

\bibitem[\protect\citeauthoryear{{Trabucchi}, {Wood}, {Montalb{\'a}n},
  {Marigo}, {Pastorelli}  \& {Girardi}}{{Trabucchi} et~al.}{2019}]{trabucchi19}
{Trabucchi} M.,  {Wood} P.~R.,  {Montalb{\'a}n} J.,  {Marigo} P.,  {Pastorelli}
  G.,   {Girardi} L.,  2019, \mn@doi [\mnras] {10.1093/mnras/sty2745}, \href
  {https://ui.adsabs.harvard.edu/abs/2019MNRAS.482..929T} {482, 929}

\bibitem[\protect\citeauthoryear{{Vauclair} et~al.,}{{Vauclair}
  et~al.}{2011}]{vauclair2011}
{Vauclair} G.,  et~al., 2011, \mn@doi [\aap] {10.1051/0004-6361/201014457},
  \href {https://ui.adsabs.harvard.edu/abs/2011A&A...528A...5V} {528, A5}

\bibitem[\protect\citeauthoryear{{Winget}, {Hansen}  \& {van Horn}}{{Winget}
  et~al.}{1983}]{winget83}
{Winget} D.~E.,  {Hansen} C.~J.,   {van Horn} H.~M.,  1983, \mn@doi [\nat]
  {10.1038/303781a0}, \href
  {https://ui.adsabs.harvard.edu/abs/1983Natur.303..781W} {303, 781}

\bibitem[\protect\citeauthoryear{{Winget} et~al.,}{{Winget}
  et~al.}{1994}]{winget94}
{Winget} D.~E.,  et~al., 1994, \mn@doi [\apj] {10.1086/174455}, \href
  {https://ui.adsabs.harvard.edu/abs/1994ApJ...430..839W} {430, 839}

\bibitem[\protect\citeauthoryear{{Xiong}, {Deng}  \& {Zhang}}{{Xiong}
  et~al.}{2018}]{xiong18}
{Xiong} D.~R.,  {Deng} L.,   {Zhang} C.,  2018, \mn@doi [\mnras]
  {10.1093/mnras/sty2014}, \href
  {https://ui.adsabs.harvard.edu/abs/2018MNRAS.480.2698X} {480, 2698}

\bibitem[\protect\citeauthoryear{{Zahn}}{{Zahn}}{1970}]{zahn70}
{Zahn} J.~P.,  1970, \aap, \href
  {https://ui.adsabs.harvard.edu/abs/1970A&A.....4..452Z} {4, 452}

\bibitem[\protect\citeauthoryear{{Zong}, {Charpinet}, {Fu}, {Vauclair}, {Niu}
  \& {Su}}{{Zong} et~al.}{2018}]{zong2018}
{Zong} W.,  {Charpinet} S.,  {Fu} J.-N.,  {Vauclair} G.,  {Niu} J.-S.,   {Su}
  J.,  2018, \mn@doi [\apj] {10.3847/1538-4357/aaa548}, \href
  {https://ui.adsabs.harvard.edu/abs/2018ApJ...853...98Z} {853, 98}

\makeatother
\end{thebibliography}


\end{document}